\documentclass{elsart}
\usepackage{natbib,psfig}

\newcommand{\ltsimeq}{\raisebox{-0.6ex}{$\,\stackrel 
        {\raisebox{-.2ex}{$\textstyle <$}}{\sim}\,$}} 
\newcommand{\gtsimeq}{\raisebox{-0.6ex}{$\,\stackrel 
        {\raisebox{-.2ex}{$\textstyle >$}}{\sim}\,$}} 

\begin{document}
\runauthor{Rawlings}
\begin{frontmatter}
\title{High-redshift radio galaxies: at the crossroads}
\author[steve]{Steve Rawlings}

\address[steve]{Astrophysics, Oxford University, 
Keble Road, Oxford, OX1 3RH, UK}
\begin{abstract}
The next generation of surveys of extragalactic radio sources will
be dominated by different types of objects than the jetted-AGN that
dominate surveys like 3C, 6C and 7C. Before radio astronomy 
becomes concentrated on the new types of object, it is
vital that we understand the cosmological importance of the 
jetted-AGN that have been studied for many years.
I argue that, as observational manifestations of
Eddington-tuned outflow events, these objects may have more 
significance for galaxy/cluster formation and
evolution than is typically appreciated. 
Outstanding problems in galaxy formation may be solved
by cosmological simulations in which this type of outflow,
as well as other types associated with weaker jets, are
properly taken into account. I will highlight areas of
ignorance which are currently hindering attempts to do this.
\end{abstract}
\end{frontmatter}

\section{Introduction}

Radio astronomers are approaching an important crossroads (Fig. 1).
The sensitivity of surveys with next-generation instruments
like LOFAR and the SKA will be such that, as
illustrated in Fig.\ 2, the radio emission from a typical survey object will 
have its origin in star formation (SF) rather than black-hole-related
processes. There will be a dramatic
increase in the areal density of `radio galaxies', and, with 
HI surveys with the SKA, direct access to 3D maps of the
large-scale structure (LSS). Radio observations may eventually lead the 
way in areas of astrophysical research traditionally dominated by
observations in other wavebands. 

This paper has a simple purpose -- to make the case that much work still
needs to be done if we are to claim to understand the jetted-AGN
that dominate existing, classical radio surveys, as
well as other types of jetted-AGN that are likely to
be common in future surveys. I will
argue that, despite their perceived rare and
exotic nature, jetted-AGN may have an important 
influence on the Universe, and must therefore be included in
cosmological simulations. I will highlight some important, but as yet 
unresolved, questions concerning these seemingly well-studied objects.

\begin{figure}
\psfig{file=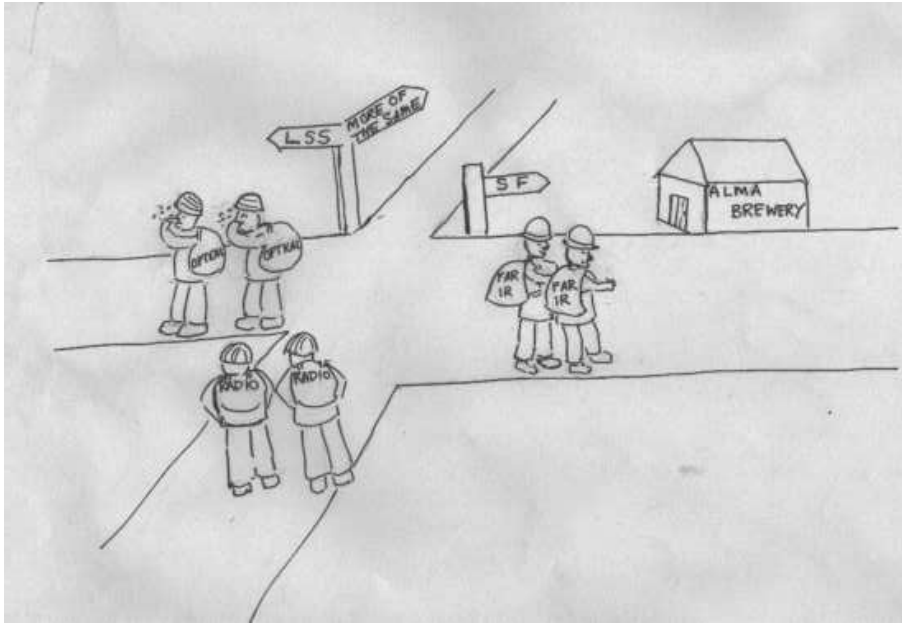,angle=0,width=12cm}
\caption{Artist's$^{\dag}$ impression of the future of 
extragalactic radio astronomy.
}
\label{fig:fig1}
\end{figure}

\begin{figure}
\psfig{file=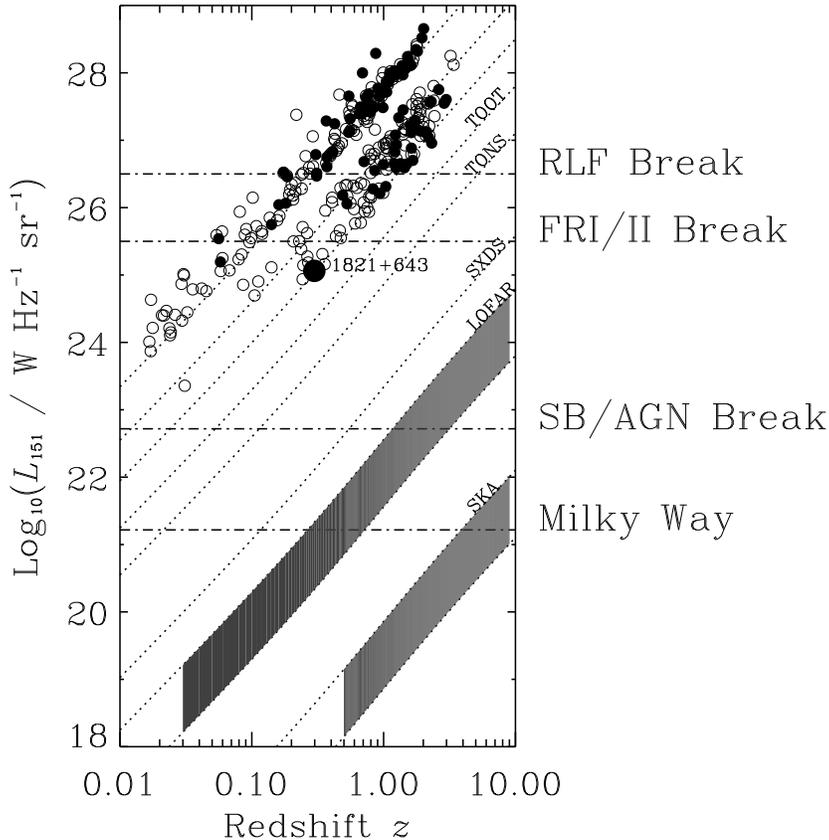,angle=0,width=12cm}
\caption{ 
151-MHz luminosity $L_{151}$ 
versus redshift $z$ diagram for various existing radio source 
redshift surveys: filled symbols are radio quasars, and open
symbols are radio galaxies from  
3C, 6CE and 7CRS; other dotted lines show loci of the
flux-density limits of TOOT \citep{HR}; TONS \citep{BRHL};
radio follow-up of the
Subaru/XMM-Newton Deep Survey (SXDS; Simpson et al., in prep.). QSOs like 
$1821+643$ are also found in the `intermediate-$L_{151}$'
zone between the plotted radio quasars and radio-quiet quasars
which tend to have detections or limits in between the 
TONS and SXDS lines.
Also plotted are bands representing 
approximate limits (depending on issues like effective exposure time
and confusion limits) for surveys with LOFAR and the SKA. Various
critical values of $L_{151}$ are marked: the break in
the RLF \citep{WRBLE}; the 
FRI/FRII break; the starburst (SB) / AGN break and the
radio luminosity of the Milky Way. Because the RLF is steep,
a typical `high-$z$ radio galaxy' in a LOFAR survey will
be a starburst rather than an AGN, and,
in SKA surveys, it will be a normal galaxy rather than a starburst.
}
\label{fig:fig2}
\end{figure}

\section{The relevance of radio sources}

Some astronomers have been heard to argue that extragalactic
radio sources are irrelevant to the energy budget of the baryons in the
Universe. Such arguments are commonly based on two
misconceptions: (i) that radio luminosity gives a reliable measure of the 
`power' required to generate radio emission; and (ii) that all power
outputs associated with AGN couple with comparable efficiency to the 
baryons. The first misconception is typified by the 
use of a sharp drop in the spectral energy distribution (SED; the plot of
$\nu \, L_{\nu}$ versus $\nu$) of AGN at radio wavebands to argue that 
radio emission is energetically unimportant. In fact, the 
bolometric radio luminosity of a radio source
is typically $\ltsimeq 1$ per cent of 
the bulk kinetic
power in its jets\footnotemark \citep{WRBL}. If one
could plot the SED of a radio source including both mechanical and 
radiative outputs, then the mechanical output might well dominate.
The second misconception arises because radiative outputs tend to be
far more obvious than mechanical outputs. However, the fact that we see
QSO radiative output so easily, e.g. 
to very large cosmological
distances, implies that radiative
outputs are often rather inefficiently absorbed by the baryons in the Universe.

\footnotetext{
All statements in this paper concerning the jet powers of radio sources
ignore the considerable uncertainties associated 
with issues such as jet composition \citep{WRBL}.
}

Having established that the mechanical output of radio sources
must be considered, it is also worth emphasising that making
any estimate of the heating effect of the radio source population
is observationally difficult. The radio luminosity 
function (RLF) has a pronounced break, and the 
luminosity density, and hence the total
heating effect of the population, is dominated by objects at or near that 
break. We see from Fig. 2 that it is only the
most recent generation of radio source redshift surveys, like TOOT
\citep{HR}, that are sensitive to sources near the break 
in the younger high-$z$ Universe, where the heating effect probably peaked
[Fig.\ 3 of \citep{R}]. Such sources can be thought of as being `typical'
of powerful-jetted-AGN in the same, luminosity-weighted, sense that 
$L^{*}$ galaxies are typical of the normal galaxy population.

Some astronomers have also been heard to argue that radio sources
are irrelevant to galaxy formation. The basis of this argument, is that
only a tiny fraction of galaxies are, at any cosmic epoch, powerful radio
sources. The hole in this argument is that the low space density of
radio-selected objects needs to be corrected for the finite and 
short ($\ltsimeq 10^{8} ~ \rm yr$) lifetimes of radio outbursts.
Accounting for this it seems likely that 
a large fraction, and possibly all, of the most massive 
($\gtsimeq 3 L_{*}$) galaxies have developed powerful radio jets on
one or more occasions during their history \citep{WRB}.

So, my contention is that radio sources may be an important, and
an as yet missing, ingredient in `semi-analytic' models 
[e.g. \citep{SP}] for 
galaxy formation. Powerful radio sources are `Eddington-tuned'
engines \citep{R} with jet powers $Q \sim f\,L_{\rm Edd}$,
with $f \sim 0.1$ \citep{WRBL}, 
and TOOT is beginning to tell us \citep{HR,RHW} that 
this is true for objects dominating the luminosity density, and hence 
the integrated heating effect, of the population at high redshift.

\section{What problems might radio sources help solve?}

One example of a problem potentially solved by considering 
radio sources is the rough calculation \citep{R} that, although most of 
the thermal energy in 
cluster baryons came from gravitational collapse, powerful
radio sources may have injected heat at the $\sim 10$ per cent level. This
may go at least some way towards explaining the `excess' entropy
inferred in the central regions of clusters \citep{PCN}.

Including radio sources in cosmological simulations
may solve other problems. The tight relation between black hole
mass and stellar luminosity, and/or stellar velocity dispersion \citep{McL},
is nicely explained by AGN feedback \citep{Silk,F}. According to feedback
models, it is the coupling of the mechanical power of AGN 
to the baryonic component of the forming galaxy 
that explains the normalization and slope of these relations.
In essence these models balance the mechanical power, delivered
presumably by winds and/or jets at some unknown fraction $f_{\rm mech}$
of the Eddington luminosity $L_{\rm Edd}$, with the power required to
push out a shell of gas at a speed $V_{\rm shell}$. So,

$$
f_{\rm mech} L_{\rm Edd} \sim V_{\rm shell} (\rho V_{\rm shell}^2) 4 \pi
r^2
\propto \frac{\sigma^2}{G r^2} r^2 V_{\rm shell}^3, 
$$

where the proportionality follows from the assumption that the 
gas density $\rho$ follows the density profile of the 
dark matter, characterized 
by a velocity dispersion $\sigma$. If the shell is to be pushed 
away from the galaxy then $V_{\rm shell}$ must exceed
$\sigma$, yielding a
relation close to that seen in the data \citep{McL}. Including 
rough estimates of the normalization,
the relation can be written as 

\begin{equation}
\left( \frac{f_{\rm mech}}{0.1} \right) \left(
\frac{L_{\rm Edd}}{10^{40} ~ \rm W} \right) 
\sim k_{1} \left( \frac{\sigma}{1000 ~ \rm km ~ s^{-1}} \right)^5,
\end{equation}

or alternatively as

\begin{equation}
\left ( \frac{f_{\rm mech}}{0.0001} \right) \left(
\frac{L_{\rm Edd}}{10^{40} ~ \rm W} \right) 
\sim k_{2} \left( \frac{\sigma}{250 ~ \rm km ~ s^{-1}} \right)^5,
\end{equation}

where $k_{1}$ and $k_{2}$ are constants of order unity, 
accounting for the numerous uncertainties.

The implication is that Eddington-tuned outflows may be more than
just `fireworks' -- they may be an integral part of galaxy formation. 
Despite at least one astronomer at the meeting saying they never wanted 
to see the phrase again, I stubbornly suggest we start thinking about
powerful radio sources less as discrete objects and
more as {\it Eddington-tuned feedback events}.
Such events may be ubiquitous in the life histories of the most
massive ellipticals
and their associated clusters. Since nearly all
clusters of galaxies contain one or more ultramassive ($\gtsimeq 3
L^{*}$) elliptical galaxy, it may be impossible to
fully understand the properties of clusters, and the elliptical galaxies
they contain, without understanding the effects of these dramatic, short-lived 
events.

\section{Radio sources as Eddington-tuned feedback events}

AGNs almost certainly drive more than one type of outflow. In fact
theories predict at least two types: (i) the narrow-beam, relativistic
radio jets we all know and love, for which the 
power source is often assumed to be rotational
energy stored in the ergosphere of the 
black hole 
as the result of previous episodes of accretion
\citep{C}; and (ii) wider-beam, sub-relativistic
winds driven off the accretion disk, for which the power source
is more directly linked to the release of 
gravitational potential energy \citep{Blan}. 
These fundamentally different processes seem likely to have
widely different values of $f_{\rm mech}$, with 
theories suggesting a significantly higher value in the case
in which the outflow is a powerful, narrow jet. 

Binary black hole merger becomes a possibility for the
$\gtsimeq 3 L_{*}$ products of the mergers that
are likely to form the most massive elliptical galaxies \citep{C}. 
This provides an
attractive mechanism both for heating the stellar core via orbital 
decay \citep{Fab}, and for the formation of a single spinning 
supermassive black hole \citep{WC}. The subsequent emergence of
a powerful jet would be expected to yield a 
step-increase in $f_{\rm mech}$. This could
drive a dramatic feedback event in which the gas is expelled
on the timescale of the radio source expansion, terminating black hole growth
as well as star formation in the circum-nuclear regions. This would be in line
with suggestions that powerful jet outflows come
{\em at the end of} significant periods of black hole growth
and associated star formation \citep{WRAD,RWHADH,J03}. 
Massive elliptical galaxies
probably underwent this formative 
phase in the young ($z \gtsimeq 4$) Universe 
when the gas could have been given more energy than its binding
energy within the associated
dark-matter halo (c.f.\ Equation 1). 
This gas, enriched by the short-lived, merger-driven burst
of star formation, could be driven to large ($\sim \rm Mpc$)
radii, where it might form an intracluster medium. 
One interesting, but highly speculative, 
possibility is that the fast clear out of recently
enriched material may be important in explaining the puzzling 
metal, i.e. [Mg/Fe], abundances of elliptical galaxies \citep{Tho}.

This all sounds reasonable. Physical models suggest
that $\approx \frac{1}{2} \sim 1$ of the jet power of radio galaxies 
finds its way into the gaseous surroundings whether it is
mediated by bubbles, at low $Q$ \citep{BK},  
or strong bow shocks, at high $Q$ \citep{BRW}.
Comparisons of $Q$ with the bolometric (radiative) luminosity
$L_{\rm Bol}$ suggest that for powerful sources
$Q \sim L_{\rm Bol} \sim 0.1 ~ L_{\rm Edd}$
\citep{WRBL}, so given a roughly unit coupling efficiency to the 
gaseous environment, we get $f_{\rm mech} \sim 0.1$
for Eddington-tuned radio jets. Any shocks driven by the 
radio sources will have radiative luminosities much less than $Q$
because of the long cooling times of the low-density material passing
through the shocks. 
From Equation 1, we see that Eddington-tuned
jets from $\gtsimeq 10^{9} ~ \rm M_{\odot}$ black holes 
have sufficient power to
expel all the gas from a proto-cluster-like potential.

There is much observational evidence in support of the notion that  
powerful radio sources should be viewed as Eddington-tuned feedback events,
important in the history of some, arguably all, 
the most massive elliptical galaxies.

\begin{itemize}

\item
Although subtle cosmic-evolution and black-hole-mass effects
are present \citep{WRJB}, the main message from the 
remarkably tight relation between stellar luminosity and
redshift (the $K-z$ relation) for radio galaxies
remains that a massive ($\gtsimeq 3 L_{*}$) elliptical galaxy seems 
a necessary prerequisite for powerful-jet production.

\item
Evidence for the importance of halo-halo mergers in the 
production of powerful jets is growing. There are 
now several clear examples [e.g. \citep{SR}] of powerful outflows 
being triggered by interpenetrating collisions of galaxies, yielding
accretion onto pairs of black holes; in other cases of AGN activity, these
black holes may have already coalesced.
There are hints that these collisions are orchestrated by the 
collapse of larger-scale structures which, on cluster scales,
may be viewed as double, merging X-ray sources \citep{Craw}
\footnotemark and which on supercluster scales may give rise to 
conspicuous aggregations of radio sources \citep{BRHL}.

\item
Evidence that a powerful radio outbursts terminate star formation
comes from anti-correlations between various quantities and the 
linear sizes of radio sources \citep{Bak,WRAD,vano,J03}. Taking linear size to be 
related to the time since jet triggering, sub-mm luminosity
(which traces heated dust), nuclear reddening (which traces all dust)
UV absorption (which traces ionized gas) and HI absorption 
(which traces neutral gas) are only high in the most recently-triggered
systems. Together, these results make a compelling case that radio source
expansion is clearing the central regions of the gas and dust needed
to fuel starbursts.

\item
Evidence for significant heat input 
by radio sources into their environments is also
mounting. A bow shock is readily visible in the {\sl CHANDRA}
image of Cygnus A \citep{Sea} but it is a minor contributor to the 
total X-ray luminosity.
In higher-$z$ systems, the expectation is that 
bow shocks will radiate a greater fraction of the energy they 
receive, largely because the densities in collapsing systems
will be higher [by a factor $\sim (1+z)^3$], and the cooling times shorter.
Such strongly
radiative bow shocks, totalling $\gtsimeq 10^{38} \rm W$, may
well have been detected at high redshift
both directly as extended soft X-ray emission
\citep{Car} and indirectly as the source of photoionizing 
photons for low-density gas in and around the lobes of $\sim 100$-kpc-scale 
radio sources \citep{Ins}.

\end{itemize}

\footnotetext{
At this meeting, Crawford argued that the latest X-ray data on
the high-$z$ radio galaxy 3C294 support an
inverse-Compton rather than thermal origin for its double X-ray structure. 
Simpson \& Rawlings (2002) have already considered this possibility,
and argued that confinement of the CMB-up-scattering
electrons would still require an underlying double
structure in the gas and dark-matter distributions.}

\section{Other types of Eddington-tuned feedback event?}

Initial results from the TOOT survey \citep{RHW} suggest that
typical high-redshift radio sources appear to be Eddington-tuned
outburst events. However, not all Eddington-tuned events,
for example `radio-quiet' quasars, seem to produce powerful jets.
This begs the question of whether jets are important only for
a sub-class of galaxies, e.g. the most massive ($\gtsimeq 3 L_{*}$)
ones. This question is linked to two old results: (i)
that some sort of 
jet activity is common, possibly ubiquitous, in radio-quiet QSOs
\citep{MRS}; and (ii) that radio-quiet quasars 
are known to reside in both spiral and elliptical galaxies \citep{McL}.

The idea of Miller, Rawlings \& Saunders (1993) --
that all radio-quiet QSOs have low-$Q$,
but initially relativistic, jets -- seemed attractive
as QSOs with intermediate radio luminosity could be explained
as the favourably aligned, Doppler-boosted cases. The
case of the quasar $1821+643$, whose credentials as 
an intermediate-$L_{151}$ quasars are clear from Fig.\ 2, has shown 
that the real situation is more complicated. Deep radio observations
of $1821+643$ \citep{BR} reveal a radio structure akin to those of classical 
FRI radio galaxies, i.e. intrinsically weak non-Doppler-boosted 
jets. Blundell \& Rawlings (2001) argue that similar
low-radio-surface-brightness features may be present in a
significant fraction of optically-selected QSOs, but they have
not yet been properly looked for.
Large-scale jets may be a far more
common feature of QSO activity than is commonly recognised.

I believe, however, that a simple but powerful argument limits the
power in such jets, and any associated winds, 
to a small fraction of the 
bolometric/Eddington luminosity of their associated 
highly accreting AGN. The
radio source population can account nicely for the 
excess entropy seen in the cores of clusters \citep{R}. 
Radio sources dominate the heating budget (over `radio-quiet'
AGN) provided that they have typical ratios of jet power 
to bolometric luminosity that exceed the factor $\sim 100$ by which 
the (bolometric) luminosity density of the radio-quiet population 
exceeds that of the radio-loud population \citep{R}. This is in accord with 
the normal assumption \citep{MRS}
that, to zeroth order, the bolometric output in jets scales
with the total low-frequency radio luminosity.
Reversing this argument, if the `radio-quiet' 
population were postulated 
to have winds (or indeed jets) 
with similar $f_{\rm mech}$ to the
radio-loud population, then the wind/jet input to the 
intracluster medium would be $\sim 100$-times greater 
than is observed in the form of the `excess' entropy in clusters.
I conclude that both jets and winds from Eddington-tuned
radio-quiet quasars have 
$f_{\rm mech} \sim 0.0001$, and hence are intrinsically weak. 

It is important to note, however, that weak jets and winds 
can have very important effects. There is 
already compelling evidence that 
fairly weak jets can limit the cooling of gas in present-day rich clusters
\citep{BK}, although in these cases the outflows are driven by
AGN accreting at sub-Eddington rates. 
AGN-driven winds from highly accreting systems could lead to
a different type of feedback event,
tuned to a lower fraction of $L_{\rm Edd}$, but still capable of 
removing gas from the central regions of the starburst/AGN,
limiting black hole growth and star formation. 
From Equation~2, material can clearly be pushed out of
the central regions of a galaxy even if it cannot reach
the scales of the intracluster medium.
This less-extreme
mode of feedback may be ubiquitous during the formation of 
less massive ($\ltsimeq 3 L^{*}$) spheroids. 
Since more massive spheroids are 
likely to grow hierarchically, their progenitors probably underwent
such events. The fact that the so-called `radio-loud' BALQSOs 
\citep{Beck} also inhabit the 
intermediate-$L_{151}$ 
zone (around $1821+643$ in Fig. 2) is very interesting in this
regard. We (Grimes et al.\ in prep) are using sub-mm and
radio observations to investigate the possibility that weak
outflows are capable of disrupting star formation like their
powerful-jet counterparts \citep{WRAD}.

\section{The hidden variable -- black-hole spin?}

\begin{figure}
\psfig{file=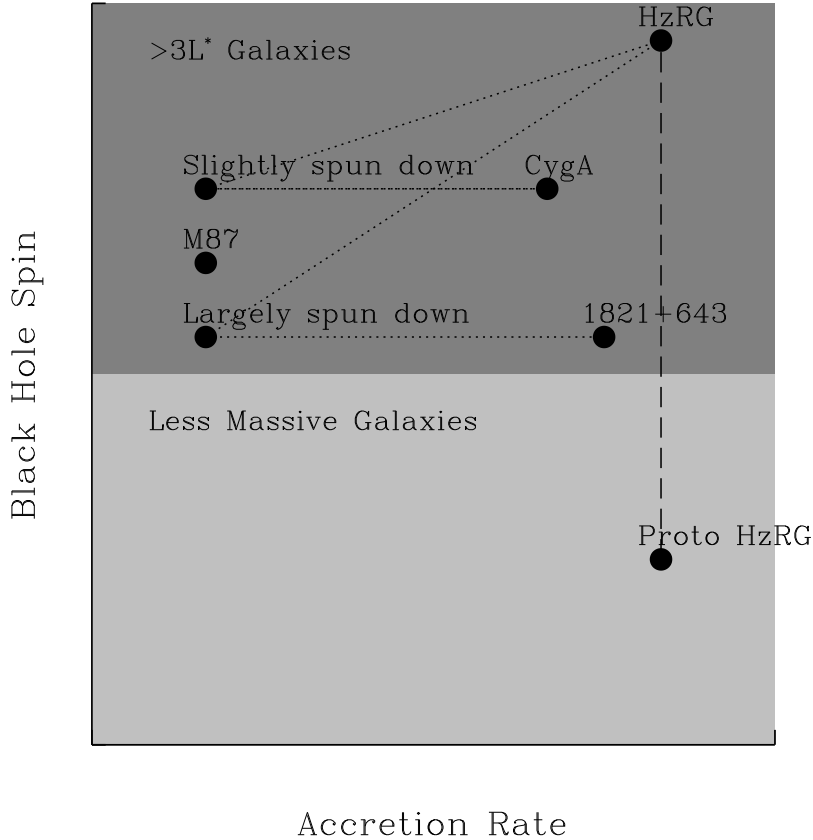,angle=0,width=12cm}
\caption{\label{fig:fig3} Sketch of possible evolutionary tracks in
the black hole spin versus accretion rate plane. Eddington-tuned
events during and after formation could spin-down the black hole
either `slightly' or `largely'. Objects similar to observed high-$z$
radio galaxies (HzRG) were likely to be the young-Universe
counterparts of diverse present-day objects like M87, Cygnus A and
$1821+643$.  Hierarchical models predict that the progenitors of
 $>3 L^{*}$ `radio loud' galaxies in the upper part of the
Figure were two less massive galaxies (probably with two black holes
and little spin) which merged at very high redshift.  }
\end{figure}

I sketch in Fig.\ 3 some possible evolutionary tracks for 
the most massive (elliptical) galaxies after their formation.
As black-hole mass is essentially a
constant along these tracks, it is clear that variations in 
black-hole mass and/or accretion rate cannot determine the wide 
variation in jet and
accretion properties. This is exemplified by the cases of M87 
(a sub-Eddington accretor with weak jets), Cygnus A
(a much-closer-to-Eddington accretor with powerful jets)
and $1821+643$ (a close-to-Eddington accretor with 
weak jets). There must be a hidden variable, and this is 
normally ascribed to black-hole spin. Very crudely, this 
might work as follows. At some suitably high redshift
(say $z \sim 7$), the first $3 L^{*}$ galaxies formed
from galaxy-size halo mergers, and they were born with
spinning supermassive black holes.
Some, but not all, of this spin was removed during the first Eddington-tuned
feedback event. Orchestrated perhaps by the collapse
of larger-scale structures, subsequent galaxy-size halo mergers 
(say at $z \sim 1-7$), and further Eddington-tuned
feedback events could have extracted more and more of spin.
So, a $\gtsimeq 3 L^{*}$ galaxy that was born in a rare dark-matter fluctuation
may well have had such a turbulent history that by present times its 
black hole has
little spin remaining. Now, situated perhaps in the
core of a very rich, relaxed cluster like $1821+643$ \citep{LRH}, 
even when its black hole is strongly accreting, it has
too little spin to power a powerful-jet outflow.
On the other hand, a $\gtsimeq 3 L^{*}$ galaxy that was
born within a less rare dark-matter fluctuation 
may well have lost little spin since its formation and
its initial Eddington-tuned feedback event. Its large-scale environment
might be dynamically young in the sense that it is part 
of a galaxy group which is only at present times
falling into a massive relaxed cluster, 
such as appears to be the case for Cygnus A \citep{OLMH}. Even moderate 
accretion
onto such a black hole might stimulate a powerful radio jet. 

This is certainly not the first time a plot such has Fig.\ 3
has been shown, and it just as
certainly won't be the last. My plea here is that
we finally, collectively, make some observational progress on understanding
it!
The crude evolutionary scenario discussed above, and its potential
relation to dark-matter fluctuations, is undoubtedly
far too simplistic but it at least provides a framework which is 
testable via observations.

\section{Concluding Remarks}

\begin{itemize}

\item 
Let's finish the job of understanding the {\em
Eddington-tuned outburst events} that are observationally
manifested as powerful
radio galaxies. I have argued here that,
when viewed at high redshift, these events may
well be the terminal phase in the formation of $\gtsimeq 3 L^{*}$
elliptical galaxies. The triggering of powerful jets 
may well introduce a step-function change in $f_{\rm mech}$
which delivers sufficient power to the environment that it influences
the gas content, and hence star-formation history, of entire 
clusters (Equation~1). The fantastic results presented 
at this meeting identifying high-$z$ radio galaxies with
forming proto-clusters \citep{Ven} is obviously in line with this 
sort of picture.

\item 
Let's finish the job of understanding the weaker jets, noting that
we need urgently to determine why these are sometimes associated with
sub-Eddington accretors (classical FRI radio galaxies) and sometimes
with highly-accreting objects (`radio-quiet' quasars like
$1821+643$). This is vital because recent results suggest that 
even weak outflows can have important influences
on the Universe, such as regulating cooling flows in
galaxy clusters. Weak, but relativistic (i.e.\ jet-like)
outflows may be just part of a 
second type of Eddington-tuned outburst event that may be the
terminating feedback process during the formation of less massive
($\ltsimeq 3 L^{*}$) spheroids. If there are associated wide-beam,
sub-relativistic (i.e.\ wind-like) outflows, I have argued that
they must carry at least a factor $\sim 100$-times less mechanical power
per unit bolometric power than the powerful radio jets. 
This is still sufficient to remove gas from the central
regions (Equation~2). We also need
to understand how such outflows are linked to potential
observational manifestations like the BALQSO phenomenon.

\item
Let's devise methods for finally hunting down the `hidden' variable in
physical models for AGN which is normally, without much observational
evidence, ascribed to black hole spin. Clues to this may come from 
studying the exact link between AGN triggering, star formation and the
collapse of dark-matter fluctuations.

\item Jetted-AGN, be they powerful-jetted-AGN
like known high-redshift radio galaxies or weak-jetted AGN, may have
much more cosmological importance than is typically recognised.
They should not, therefore, be treated as
just a nuisance for future surveys with LOFAR and the SKA.

\end{itemize}

\section*{Acknowledgements}
I thank all my collaborators and the organizers of
what proved to be a great meeting. Special
thanks to Matt Jarvis and the artist$^{\dag}$, my mum, responsible
for Fig. 1.

\end{document}